\documentclass[lettersize,journal]{IEEEtran}

\usepackage{cite}
\usepackage{amsmath,amssymb,amsfonts}
\usepackage{algorithmic}
\usepackage{graphicx}
\usepackage{textcomp}
\usepackage{xcolor}

\usepackage{bm, amssymb, booktabs, multirow,cite, url, xcolor, hyperref}
\usepackage{makecell, arydshln}
\usepackage{threeparttable}

\usepackage[T1]{fontenc}
\usepackage[utf8]{inputenc}
\usepackage{verbatim}

\definecolor{revblue}{RGB}{0,70,140}
\definecolor{revgreen}{RGB}{0,120,80}
\definecolor{revred}{RGB}{150,40,40}

\newif\ifshowrev
\showrevtrue          % show colored markers
\showrevfalse       % hide markers

\newcommand{\reviewerone}[1]{\ifshowrev{\color{blue}#1}\else#1\fi}

\newcommand{\reviewertwo}[1]{\ifshowrev{\color{revgreen}#1}\else#1\fi}

\def\BibTeX{{\rm B\kern-.05em{\sc i\kern-.025em b}\kern-.08em
    T\kern-.1667em\lower.7ex\hbox{E}\kern-.125emX}}
\usepackage{balance}
\begin{document}
\title{Efficient and Robust Speaker Diarization via Structured Pruning of Self-Supervised Models}
% \title{Efficient and Generalizable Speaker Diarization via Structured Pruning of Self-Supervised Models}
\author{Jiangyu Han, Petr Pálka, Marc Delcroix, Federico Landini, Johan Rohdin, Jan Černocký, Lukáš Burget 
\thanks{Jiangyu Han, Petr Pálka, Federico Landini, Johan Rohdin, Jan Černocký, and Lukáš Burget are with 
Brno University of Technology, Speech@FIT, Czechia Republic.
Marc Delcroix is with NTT, Inc., Japan. Corresponding author:  ihan@fit.vut.cz
}
}

\markboth{Journal of \LaTeX\ Class Files,~Vol.~18, No.~9, September~2020}%
{How to Use the IEEEtran \LaTeX \ Templates}

\maketitle

\begin{abstract}
\reviewerone{This work presents a framework for compressing self-supervised models for speaker diarization through structured pruning guided by knowledge distillation. We investigate pruning objectives that target reducing both model parameters and computational complexity, where knowledge distillation enables compact models and pruning removes unnecessary parameters to improve hardware efficiency. We further analyze alternative pruning strategies, showing that a simple overall pruning approach provides the best balance between efficiency and accuracy. Compared to the original unpruned model, our method achieves up to 80\% model size reduction and 4x faster inference without performance degradation.} Comprehensive experiments across eight public diarization datasets demonstrate that the pruned models consistently match or surpass the performance of their uncompressed counterparts. Furthermore, we show strong out-of-domain generalization on the CHiME-6 dataset, achieving accuracy comparable to the top systems in the CHiME-7 challenge without any domain adaptation. These results highlight that structured pruning, when guided by distillation, can yield efficient and generalizable diarization systems suitable for real-world applications.

\end{abstract}


\vspace{+0.1cm}

\section{Introduction}
Understanding multi-speaker conversations requires identifying when each person is speaking and attributing speech segments to the correct speaker—a task known as \textit{speaker diarization}. 
In recent years, pre-trained self-supervised learning (SSL) models have revolutionized speech representation learning, leading to major advances across a wide range of tasks~\cite{mohamed2022self}. 
Among them, WavLM~\cite{chen2022wavlm} has become a cornerstone in state-of-the-art speaker diarization systems~\cite{tawara2024ntt, plaquet2024mambabasedsegmentationmodelspeaker, han2025leveraging}. 
These systems leverage the contextualized acoustic representations offered by SSL models, substantially reducing diarization error rates (DER) in complex conversational scenarios. 
Despite these gains, SSL-based diarization systems remain difficult to deploy in practice due to the high computational and memory demands of large models such as WavLM. 
With hundreds of millions of parameters, these models incur high inference latency and require considerable hardware resources, limiting their use in latency-critical and on-device applications.

The limitations of large pre-trained models have motivated growing interest in model compression techniques~\cite{cheng2024survey} aimed at reducing resource consumption while preserving performance. 
Two complementary approaches have been widely explored for this purpose. 
\textit{Knowledge distillation} trains a compact student to imitate a larger teacher~\cite{chang2022distilhubert, ashihara2022deep, gandhi2023distil}, while \textit{structured pruning}~\cite{louizos2018learning, wang2019structured, xia2022structured, xia2023sheared, athanasoulias2025isomorphic} automatically removes groups of redundant parameters—such as attention heads or intermediate dimensions of feed-forward networks—from large pre-trained models. 
Unlike unstructured pruning, which removes individual weights and typically depends on support for sparse matrix operations~\cite{sanh2020movement, huang2021sparse}, structured pruning preserves architectural regularity, making it more compatible with standard inference frameworks and capable of delivering real improvements in runtime and memory efficiency. 
Moreover, combining pruning with distillation objectives has been shown to further enhance the performance of compressed models~\cite{sanh2020movement, lagunas2021block, xia2022structured}.

In the field of speech processing, model compression has been extensively explored for SSL models through the SUPERB benchmark~\cite{peng2023dphubert, chang2022distilhubert} and in automatic speech recognition (ASR) tasks~\cite{peng2023structured, jiang2023accurate}. 
Although their results are promising, performance often degrades sharply at high pruning ratios. 
Speaker diarization poses an additional challenge, as it must handle long multi-speaker recordings under diverse acoustic conditions—yet it may require less representational capacity than content-based tasks such as ASR. 
This raises an important question: \textit{can large self-supervised models be pruned more aggressively for diarization without loss of accuracy?} 
In this paper, we show that the answer is yes: structured pruning guided by distillation can yield compact and efficient diarization models that generalize well across domains.

\reviewertwo{In our previous work~\cite{han2025fine}, we investigated structured pruning of WavLM models for speaker diarization. 
Unlike existing approaches that often struggle to maintain acoustic representation integrity at high sparsity, we demonstrated that fine-tuning the WavLM model with a diarization objective before structured pruning is instrumental in guiding self-supervised models to retain robust task-specific knowledge during the compression process. With this strategy, we achieved up to 80\% parameter reduction for both WavLM Base+ and WavLM Large models, together with 4.0$\times$ and 2.6$\times$ inference speedups on a single GPU without performance loss. Nevertheless, a comprehensive analysis of structured pruning behaviors was beyond the scope of that study.}

\reviewertwo{Building on these initial findings, in this paper, we} \reviewerone{incorporate} new pruning objectives based on the number of multiply-accumulate operations (MACs), alternative pruning strategies, and a substantially broader experimental evaluation into the framework.
We also analyze the influence of training data quantity on pruning effectiveness, showing that using only half of the available data can substantially reduce training time while maintaining comparable performance. 
Whereas our earlier study focused on meeting-style datasets such as AMI~\cite{carletta2005ami, kraaij2005ami}, AISHELL-4~\cite{fu2021aishell}, and AliMeeting~\cite{yu2022m2met}, the present work expands to a diverse compound dataset covering eight public diarization corpora—AMI, AISHELL-4, AliMeeting, NOTSOFAR-1~\cite{vinnikov2024notsofar}, MSDWild~\cite{liu2022msdwild}, DIHARD-3~\cite{ryant2020third}, RAMC~\cite{yang2022open}, and VoxConverse~\cite{chung2020spot}—and further evaluates generalization to out-of-domain data using CHiME-6~\cite{watanabe2020chime}. 
Notably, even without any domain adaptation, our method remains competitive and achieves results comparable to the third-place system in the CHiME-7 challenge~\cite{cornell2023chime}.

% Building on our previous work~\cite{han2025fine}, which achieved up to 80\% parameter reduction for both WavLM Base+ and WavLM Large models with 4.0$\times$ and 2.6$\times$ inference speedups on a single GPU, we \reviewerone{incorporate} new pruning objectives based on the number of multiply-accumulate operations (MACs), alternative pruning strategies, and a substantially broader experimental evaluation into the framework.
% We also analyze the influence of training data quantity on pruning effectiveness, showing that using only half of the available data can substantially reduce training time while maintaining comparable performance. 
% Whereas our earlier study focused on meeting-style datasets such as AMI~\cite{carletta2005ami, kraaij2005ami}, AISHELL-4~\cite{fu2021aishell}, and AliMeeting~\cite{yu2022m2met}, the present work expands to a diverse compound dataset covering eight public diarization corpora—AMI, AISHELL-4, AliMeeting, NOTSOFAR-1~\cite{vinnikov2024notsofar}, MSDWild~\cite{liu2022msdwild}, DIHARD-3~\cite{ryant2020third}, RAMC~\cite{yang2022open}, and VoxConverse~\cite{chung2020spot}—and further evaluates generalization to out-of-domain data using CHiME-6~\cite{watanabe2020chime}. 
% Notably, even without any domain adaptation, our method remains competitive and achieves results comparable to the third-place system in the CHiME-7 challenge~\cite{cornell2023chime}.
The main contributions of this paper are as follows:
\begin{enumerate}
    % \item We extend our structured pruning framework with new pruning objectives, sparsity scheduling strategies, and data-efficient training schemes.
    \item \reviewerone{We incorporate new pruning objectives, sparsity scheduling strategies, and data-efficient training schemes into our structured pruning framework.}
    \item We conduct comprehensive evaluations across eight diverse diarization corpora and demonstrate robust out-of-domain generalization without domain adaptation. Notably, our pruned models achieve state-of-the-art performance on the majority of public benchmark datasets.
    \item We release all trained models and code to support reproducibility and future research within the speaker diarization community\footnote{\url{https://github.com/BUTSpeechFIT/DiariZen}}.
\end{enumerate}

\section{\reviewerone{Knowledge Distillation-Guided Structured Pruning Framework}}

\begin{figure}[tbp]
  \centering
  \includegraphics[width=8cm]{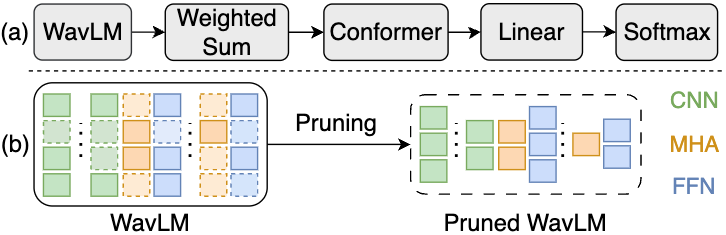}
  \caption{ 
 Overview of the proposed method: (a) WavLM-based end-to-end diarization model used for fine-tuning, 
  and (b) structured pruning applied to its WavLM component (light-colored regions denote pruned units).
 }
  \label{fig:framework}
\end{figure}

The proposed framework follows a three-stage pipeline to obtain compact yet high-performing diarization models from large self-supervised backbones.  
\textbf{Stage~1:} A pre-trained WavLM model is first fine-tuned within an end-to-end diarization framework to adapt it to the speaker diarization task.  
\textbf{Stage~2:} The fine-tuned model then undergoes \emph{structured pruning} guided by \emph{knowledge distillation}, which removes redundant components while preserving the performance of the task. 
\textbf{Stage~3:} The pruned model is re-finetuned within the same diarization pipeline to recover any performance loss.  
The overall architecture and pruning strategy are illustrated in Figure~\ref{fig:framework}, where pruning operates across different architectural components, including individual convolutional neural network (CNN) kernels, complete attention heads in multi-head attention (MHA) layers, and rows and columns of weight matrices corresponding to specific intermediate dimensions in the feed-forward (FFN) layers. Detailed training configurations for these stages are provided in Section~\ref{subsec: configurations}, and an ablation of alternative pruning strategies is presented in Section~\ref{subsub: pruning_alternatives}.

\subsection{Speaker diarization pipeline}
\reviewertwo{This subsection describes the speaker diarization pipeline used in the proposed framework, which corresponds to the model architecture employed in both \textbf{Stage~1 (pre-fine-tuning)} and \textbf{Stage~3 (re-fine-tuning after pruning)}.}  
We adopt DiariZen~\cite{han2025leveraging}, 
which follows the EEND-VC (end-to-end neural diarization with vector clustering) paradigm~\cite{kinoshita2021integrating, kinoshita2021advances} 
by combining a local end-to-end neural diarization (EEND) module 
with a Pyannote-based~\cite{bredin2023pyannote} backend for speaker clustering and stitching of local-window decisions. 
As shown in Figure~\ref{fig:framework}(a), the EEND module comprises a pre-trained WavLM~\cite{chen2022wavlm}, a Conformer~\cite{gulati2020conformer}, and a linear classifier.

\reviewertwo{Formally, given an input speech signal $\mathbf{x}$ with $T$ frames, speaker diarization aims to estimate frame-level speaker labels. 
Conventionally, it is formulated as a multi-label classification problem~\cite{fujita2019end}, where speaker activity is represented by a binary matrix $\mathbf{Y} \in \{0,1\}^{T \times C}$ indicating the joint activity of $C$ speakers. 
Alternatively, as adopted in this paper, diarization is modeled as a multi-class classification problem~\cite{plaquet2023powerset}, where each frame is assigned to one of $K$ mutually exclusive classes (e.g., silence, single-speaker activity, or overlapping speech).

}

Following SUPERB~\cite{yang2021superb}, hidden representations from all WavLM layers are combined via a learnable layer-wise weighted sum used as input to the Conformer. 
The Conformer consists of four blocks, each comprising a feed-forward module (input and hidden dimensions of 256 and 1024, respectively), a multi-head self-attention module with four heads, and a convolution module with kernel size~31. 
All dropout rates in the Conformer are set to~0.1. 
The Conformer output is passed to a linear layer that produces frame-level logits, and a final softmax yields powerset labels for the EEND objective. 
The model is trained with the powerset loss~\cite{plaquet2023powerset}, supporting up to four speakers with at most two overlapping speakers.

We consider both WavLM Base+ and WavLM Large variants (approximately 94.4\,M and 316.6\,M parameters, respectively), while the Conformer and linear layers contribute an additional 6.1\,M parameters.

\subsection{Structured pruning guided by knowledge distillation}
\reviewertwo{This subsection describes \textbf{Stage~2 (pruning + distillation)}  of the proposed framework.} 
\reviewerone{
Our objective here is to provide a self-contained description of structured pruning as adopted in our framework. 
A complete theoretical treatment is beyond the scope of this paper, and interested readers may consult the original work~\cite{louizos2018learning} for further theoretical details if necessary.
} 

As illustrated in Figure~\ref{fig:framework}(b), structured pruning is applied exclusively to the WavLM model. Following prior studies~\cite{louizos2018learning, xia2022structured, peng2023dphubert}, we formulate pruning as minimizing
\begin{equation}
\label{eq:objective}
\mathcal{L}(\bm\theta) + \lambda \Vert \bm\theta \Vert_0,
\end{equation}
where $\mathcal{L}(\bm\theta)$ is the training loss, $\Vert \bm\theta \Vert_0$ is an $L_0$ regularization term that penalizes the number of nonzero parameters (thus encouraging pruning), and $\lambda>0$ controls the pruning strength. To ensure structured pruning, we group the parameters of WavLM into $J$ disjoint groups (e.g., CNN kernels or MHA heads) and count parameters in $\Vert\bm\theta\Vert_0$ as nonzero when at least one parameter in their group is nonzero; thus, all parameters within a group are pruned or retained jointly. We therefore denote the prunable parameters as ${\bm\theta}=\{{\bm\theta}_j\}_{j=1}^J$, where  ${\bm\theta}_j$ represents the parameters of $j$th group.

\paragraph*{Distillation loss}
While various options for the training loss $\mathcal{L}(\bm\theta)$ are possible (see Sec.~\ref{subsub: pruning_alternatives}), we obtain the best performance using a distillation-based loss inspired by~\cite{peng2023dphubert}. Specifically, the fine-tuned WavLM serves as both the fixed teacher and the initialization for the student, whose parameters $\bm\theta$ are trainable and prunable. Given the $t$th frame outputs from the $i$th Transformer layer of the teacher $\bm h_t^{(i)}$ and student $\hat{\bm h}_t^{(i)}$, the distillation loss combines $L_1$ penalty and cosine distances with equal weights~\cite{chang2022distilhubert, peng2023dphubert}:
\begin{equation}
\mathcal{L}(\bm\theta)=
\sum_{i\in \mathcal{S}}\sum_{t=1}^T
L_1(\bm h_t^{(i)},\hat{\bm h}_t^{(i)})
-\cos(\bm h_t^{(i)},\hat{\bm h}_t^{(i)}),
\end{equation}
where $T$ is the number of frames and \reviewerone{$\mathcal{S}$} denotes the set of transformer layers used for teacher–student matching. 
We set \reviewerone{$\mathcal{S}=\{0,4,8,12\}$} for the Base+ model and \reviewerone{$\mathcal{S}=\{0,8,16,24\}$} for the Large model, where index $0$ denotes the input to the first transformer layer.
\reviewerone{Note that the distillation loss only constrains the prunable student model to mimic the behavior of its teacher and does not determine which modules are pruned. For example, although the distillation loss is computed on the outputs of the transformer layers, in our framework, all CNN and Transformer layers are target to pruning.}
We next describe how this loss is combined with structured pruning through stochastic gating to enforce sparsity in WavLM.

\paragraph*{Stochastic gating and relaxation}Direct optimization of~\eqref{eq:objective} is intractable due to the discrete, combinatorial, and non-differentiable nature of $\Vert\bm\theta\Vert_0$. Following~\cite{louizos2018learning}, we express each parameter group as ${\bm\theta}_j = \tilde{\bm\theta}_j z_j$, where $\tilde{\bm\theta}_j$ are learnable weights and $z_j$ is a stochastic gate indicating whether the group is active. 
Modeling the gates as independent Bernoulli random
variables and denoting their collection as $\bm z=\{z_j\}_{j=1}^J$ allows the original objective~\eqref{eq:objective}
to be equivalently rewritten as the minimization of its expected value over $\bm z$:
\begin{equation}
\label{eq:kd_l0}
\min_{\tilde{\bm\theta},\bm\alpha}
\mathbb{E}_{q(\mathbf{z}|\bm\alpha)}
\!\left[
\mathcal{L}(\bm\theta)+\lambda\Vert\bm\theta\Vert_0
\right],
\end{equation}
where $\bm\alpha=\{\alpha_j\}$ are the parameters controlling the activation probability of each gate. 
Note that $\bm\theta$ in the expectation depends on $\mathbf{z}$ through the stochastic gating mechanism defined above. 
At the optimum, the Bernoulli gates become deterministic (i.e., each $z_j$ equals 0 or 1 with probability 1), thus defining the  pattern of structured pruning.

If $z_j$ were truly Bernoulli, optimization of~\eqref{eq:kd_l0} would remain intractable because of non-differentiability. Following~\cite{louizos2018learning}, we replace the Bernoulli gates with \emph{Hard-Concrete} gates---continuous relaxations of Bernoulli distributions defined on $[0,1]$. This formulation assigns nonzero probability mass to $z_j=0$, and its location parameter $\alpha_j>0$ controls the sparsity level: smaller $\alpha_j$ yields a higher probability of $z_j$ being zero.

\paragraph*{Reparameterization}
To compute gradients of $\mathbb{E}_{q(\mathbf{z}|\bm\alpha)}[\mathcal{L}(\bm\theta)]$ 
in~\eqref{eq:kd_l0} with respect to both $\tilde{\bm\theta}$ and $\bm\alpha$, 
the expectation is approximated empirically via samples of $z_j$ drawn from the Hard-Concrete distributions using the reparameterization trick~\cite{louizos2018learning,xia2022structured}:
\begin{align}
\label{eq:reparam}
s_j &= \text{sigmoid}\!\left(\frac{\log u_j-\log(1-u_j)+\log\alpha_j}{\beta}\right), \notag\\
z_j &= \min(1,\max(0,(\zeta-\gamma)s_j+\gamma)),
\end{align}
where $u_j\!\sim\!\mathcal{U}(0,1)$, $\beta$ is a temperature parameter controlling how closely the continuous distribution approximates a discrete gate, $s_j\!\in\!(0,1)$ follows a binary-concrete distribution~\cite{jang2016categorical,maddison2016concrete}, and $(\gamma,\zeta)=(-0.1,1.1)$ stretch $s_j$ beyond $(0,1)$. The final clamping to $(0,1)$ yields a nonzero probability mass at $z_j=0$. Following~\cite{louizos2018learning,peng2023dphubert}, we fix $\beta=2/3$ throughout our experiments. Equation~\eqref{eq:reparam} thus provides a differentiable reparameterization of $z_j$ for back-propagation.

\paragraph*{Expected sparsity and optimization}
The Hard-Concrete distribution also enables a closed-form expression for the expected $\Vert\bm\theta\Vert_0$ term, allowing its gradient with respect to $\bm\alpha$ to be computed without Monte Carlo sampling:
\begin{align}
\mathbb{E}_{q(\mathbf{z}|\bm\alpha)}[\Vert\bm\theta\Vert_0]
= \sum_{j=1}^J N_j
\,\text{sigmoid}\!\left(\log\alpha_j-\beta\log\frac{-\gamma}{\zeta}\right),
\end{align}
where $N_j$ is the number of parameters in group $j$. To precisely control sparsity, we adopt the augmented-Lagrangian formulation~\cite{wang2019structured}:
\begin{align}
\label{eq:final}
\max_{\lambda_1,\lambda_2}
\min_{\tilde{\bm\theta},\bm\alpha}
\mathbb{E}_{q(\mathbf{z}|\bm\alpha)}
\!\big[
\mathcal{L}(\bm\theta)
+\lambda_1(\Vert\bm\theta\Vert_0-t)
+\lambda_2(\Vert\bm\theta\Vert_0-t)^2
\big],
\end{align}
where $\lambda_1$, $\lambda_2 \in \mathbb R$ are learnable Lagrange multipliers, initialized to zeros and dynamically updated during training,
and $t$ is the pre-defined target sparsity.

\paragraph*{Inference}
At inference time, the stochastic gates are replaced by deterministic estimates
\[
\hat z_j=\min(1,\max(0,(\zeta-\gamma)
\,\text{sigmoid}(\log\alpha_j)+\gamma)),
\]
and the remaining parameters $\tilde{\bm\theta}_j$ with $\hat z_j>0$ are retained, yielding a dense and hardware-efficient pruned model suitable for deployment in real-time diarization systems.

\begin{table*}[tbp]
% \vspace{-0.5cm}
  \caption{Information about the different datasets includes the number of recordings, the minimum and maximum number of speakers, total duration (in hours), and the average length (in minutes). The characteristics of each dataset are also provided.}
  \label{tab:dataset}
 \setlength{\tabcolsep}{1.2mm}
  \centering
  \begin{tabular}{l | c c r c | c c r c | c c r c | l }
    \hline
        \multirow{2}{*}{Dataset} & \multicolumn{4}{c|}{Train} & \multicolumn{4}{c|}{Development} & \multicolumn{4}{c|}{Test} & \multirow{2}{*}{Characteristics} \\
        & \#recs & \#spk & \#hrs & min/rec & \#recs & \#spk & \#hrs & min/rec & \#recs & \#spk & \#hrs & min/rec & \\
    \hline
    AMI & 134 & 3-5 & 79.7 & 35.7 & 18 & 4 & 9.7 & 32.3 & 16 & 3-4 & 9.1 & 34.1 & Meetings in English in different rooms \\
    AISHELL-4 & 173 & 3-7 & 97.2 & 33.8 & 18 & 4-7 & 9.9 & 33.8 & 20 & 5-7 & 12.7 & 38.2 & Discussions in Mandarin in different rooms\\
    AliMeeting & 209 & 2-4 & 111.4 & 32.0 & 8 & 2-4 & 4.2 & 31.5 & 20 & 2-4 & 10.8 & 32.4 & Meetings in Mandarin in different rooms \\
    NOTSOFAR-1 & 526 & 4-8 & 54.3 & 6.2 & 117 & 4-6 & 12.2 & 6.2 & 160 & 3-7 & 16.7 & 6.3 & Meetings in English in different rooms \\
    MSDWild & 2476 & 2-4 & 66.1 & 1.6 & 177 & 3-10 & 4.1 & 1.4 & 490 & 2-4 & 9.9 & 1.2 & Daily casual conversations (multilingual) \\
    DIHARD3 & 204 & 1-10 & 27.4 & 8.1 & 50 & 1-8 & 6.7 & 8.1 & 259 & 1-9 & 33.0 & 7.7 & Wide variety of domains \\
    RAMC & 289 & 2 & 149.6 & 31.1 & 19 & 2 & 9.9 & 31.2 & 43 & 2 & 20.6 & 28.8 & Phone calls in Mandarin \\
    VoxConverse & 174 & 1-20 & 16.5 & 5.7 & 42 & 1-15 & 3.8 & 5.4 & 232 & 1-21 & 43.5 & 11.3 & Wide variety of domains (multilingual) \\
    CHiME-6 & 14 & 4 & 35.4 & 151 & 2 & 4 & 4.5 & 134 & 4 & 4 & 10.0 & 150 & Dinner parties in English at home \\  
    \hline
  \end{tabular}
  % \vspace{-0.1cm}
\end{table*}

\subsection{Structured pruning based on MACs}
\reviewertwo{This subsection extends \textbf{Stage~2 (pruning + distillation)} by introducing MAC-based structured pruning.}
In addition to parameter count, pruning can also be guided by the number of multiply–accumulate operations (MACs), providing a more direct estimate of computational cost. The key difference from parameter-based pruning lies in the sparsity objective, which is now defined with respect to the total MACs, rather than the number of parameters. Specifically, we compute the MACs for each prunable group and define the computational sparsity as the ratio of pruned to total MACs.
% \reviewerone{\begin{equation}
% \frac{\mathrm{MAC}_{\mathrm{pruned}}}
% {\mathrm{MAC}_{\mathrm{total}}}
% \end{equation}}

Following~\cite{peng2023structured}, we estimate MACs using the formulas implemented in the \emph{DeepSpeed}\footnote{\url{https://github.com/deepspeedai/DeepSpeed}} FLOPs profiler. For an input sequence of length $T$ and hidden size $d$, the MACs for the MHA and FFN modules are
\begin{align}
\mathrm{MAC}_{\mathrm{MHA}} &= 4Thd d^{\mathrm{head}} + 2T^2 h d^{\mathrm{head}}, \\
\mathrm{MAC}_{\mathrm{FFN}} &= 2T d d^{\mathrm{int}},
\end{align}
where $h$ is the number of attention heads, $d^{\mathrm{head}}$ is the dimensionality of each head, and $d^{\mathrm{int}}$ is the size of the intermediate FFN layer. 
For a 1D convolution layer with kernel size $K$, $C^{\mathrm{in}}$ input channels, and $C^{\mathrm{out}}$ output channels, the corresponding MACs are
\begin{equation}
\label{eq:mac_ffn}
\mathrm{MAC}_{\mathrm{CNN}} = T^{\mathrm{out}} C^{\mathrm{out}} C^{\mathrm{in}} K,
\end{equation}
where $T^{\mathrm{out}}$ is the length of the output sequence. 

For each prunable group, the associated MACs depend on $h$, $d$, $d^{\mathrm{int}}$, $C^{\mathrm{in}}$, and $C^{\mathrm{out}}$. Because the total MAC count is strongly correlated with input sequence length, we consistently estimate MACs using a one-second pseudo input signal throughout this paper for comparability.

\begin{table*}[htbp]
% \vspace{-0.cm}
  \caption{Diarization performance across different datasets. Inference speedups on both GPU and CPU are reported relative to the unpruned model. Evaluations were conducted on a single NVIDIA RTX A5000 GPU and an Intel Xeon E5-2640 v4 2.40 GHz CPU, respectively, with the input batch size tuned to maximize utilization of each device computational capacity.
}
  \label{tab:overall1}
  \centering
  \begin{tabular}{l| c c | c c | c c | c c c | c}
    \hline
        \multirow{2}{*}{System} &  \multicolumn{2}{c|}{Pruning} & \multicolumn{2}{c|}{Complexity} & \multicolumn{2}{c|}{Speedup$\nearrow$} & \multicolumn{3}{c|}{DER (\%)} & 
        \multirow{2}{*}{Macro} \\
        & Sparsity & Objective & Params & MACs & GPU & CPU & AMI & AISHELL-4 & AliMeeting &  \\
    \hline
    Fbank & - & - & - & - & - & - & 19.7 & 12.5 & 21.0 & 17.7 \\
    \hline
    \multirow{5}{*}{WavLM Base+} & 0\% & - & 94.4M & 6.9G & - & - & 15.6 & 11.8 & 17.7 & 15.0 \\ 
        & 80\% & Params & 18.8M & 1.1G & 4.0$\times$ & 4.5$\times$ & 15.8 & 12.1 & 17.9 & 15.3 \\
        & 80\% & MACs & 26.5M & 1.4G & 3.5$\times$ & 3.9$\times$ & 15.7 & 12.0 & 17.9 & 15.2 \\
        & 90\% & Params & 9.4M & 0.6G & 5.9$\times$ & 7.5$\times$ & 16.9 & 11.8 & 18.5 & 15.7 \\
        & 90\% & MACs & 13.3M & 0.7G & 5.4$\times$ & 6.8$\times$ & 16.5 & 12.2 & 19.2 & 16.0 \\
    \hline
    \multirow{5}{*}{WavLM Large} & 0\% & - & 316.6M & 17.8G & - & - & 14.8 & 11.3 & 16.3 & 14.1 \\ 
        & 80\% & Params & 63.3M & 3.8G& 2.6$\times$ & 3.1$\times$ & 15.1 & 11.3 & 15.8 & 14.1 \\
        & 80\% & MACs & 70.2M & 3.6G& 3.5$\times$ & 4.3$\times$ & 15.2 & 11.2 & 16.7 & 14.4 \\
        & 90\% & Params & 30.6M & 1.8G & 3.4$\times$ & 5.2$\times$ & 15.7 & 11.2 & 17.6 & 14.8 \\
        & 90\% & MACs & 35.2M & 1.8G & 5.2$\times$ & 7.2$\times$ & 15.8 & 11.2 & 17.6 & 14.8 \\
    \hline
  \end{tabular}
  % \vspace{-0.15cm}
\end{table*}

\section{Experimental Setup}
\subsection{Datasets}
\label{subsec: datasets}
For experiments in Sections \ref{subsec: overall}, \ref{subsec: cnn_trans_inference_time}, \ref{subsec: cnn_trans_separate_pruning}, \ref{subsec: analysis_different_sparsity}, and \ref{subsec: ablation_analysis}, 
we use the far-field single-channel recordings from AMI
\cite{carletta2005ami, kraaij2005ami}, AISHELL-4 \cite{fu2021aishell}, and AliMeeting \cite{yu2022m2met} for system evaluation.  Our model is trained on the combination of the training sets from these three corpora, and their corresponding development sets are merged for validation.
Then, in the Section~\ref{sec: compound_datasets}, we further evaluate the effectiveness of our method on a larger, multi-domain compound dataset, consisting of AMI, AISHELL-4, AliMeeting, NOTSOFAR-1 \cite{vinnikov2024notsofar}, MSDWild \cite{liu2022msdwild}, DIHARD-3 \cite{ryant2020third}, RAMC \cite{yang2022open}, and VoxConverse \cite{chung2020spot} corpora. Finally, as shown in Section \ref{subsec: ood_chime}, the out-of-domain performance on CHiME6 \cite{watanabe2020chime} is also considered. 

Note that for AMI and AliMeeting, we consistently use the first channel from the first far-field microphone array. For AISHELL-4, we convert the multi-channel recordings into a single-channel audio by averaging across channels.
When constructing the compound dataset, we directly use the official train/dev/test splits when available.
For AISHELL-4 and DIHARD3, which provide only training or development data, we randomly split the available data into new training and development subsets using an 80/20\% ratio, while keeping the original test sets unchanged. These subsets are then combined to form the compound dataset.
For AliMeeting, the eval set is used for development and the test set is used for system evaluation.
For NOTSOFAR-1, we follow the official single-channel track condition\footnote{\url{https://www.chimechallenge.org/challenges/chime8/task2/data}}.
For MSDWild, we adopt the official partition into
few.train/many.val/few.val as train/dev/test. 
For DIHARD3, we use the official “full” partition.
For VoxConverse, we use the latest annotations\footnote{\url{https://github.com/joonson/voxconverse/tree/master}} from Version 0.3.
For CHiME-6, we follow the data split established in the CHiME-7 challenge \cite{cornell2023chime}. 
Detailed information can be found in Table~\ref{tab:dataset}.

\subsection{Configurations}
\label{subsec: configurations}
All models are trained using the AdamW optimizer~\cite{loshchilov2018decoupled}. We now detail the training settings for our three stage pipeline.

\textbf{Stage~1 (pre-fine-tuning).}  
In the first stage, the complete EEND system (including the WavLM front end) is optimized using the powerset loss to adapt WavLM to the diarization task. The learning rate is set to $2\!\times\!10^{-5}$ for WavLM parameters and $1\!\times\!10^{-3}$ for all remaining parameters. Training stops if the validation loss fails to improve for 10~epochs, typically converging within 20~epochs.

\textbf{Stage~2 (pruning + distillation).}  
Only the WavLM parameters are pruned and updated using a knowledge distillation objective. This stage begins by gradually increasing the target sparsity over the first five epochs. The target sparsity is then kept fixed for the remainder of the pruning phase, which usually takes about 25~epochs before early stopping is triggered by the validation loss showing no improvement for 5~epochs. During this stage, the learning rate for the student WavLM parameters $\bm{\theta}$ is set to $2\!\times\!10^{-4}$, 
and the learning rates for the sparsity-related parameters $\bm{\alpha}$, $\lambda_1$, and $\lambda_2$ are set to $2\!\times\!10^{-2}$.

\emph{(distillation continuation within Stage~2)}  
Following~\cite{peng2023dphubert}, the pruning pattern is then frozen and distillation continues without further pruning for about 15~additional epochs, again stopping early if the validation distillation loss does not improve for 5~epochs. In this phase, 
the sparsity-related parameters remain fixed, and
the learning rate for the student WavLM parameters $\bm{\theta}$ is maintained at $2\!\times\!10^{-4}$.

\textbf{Stage~3 (re-fine-tuning after pruning).}  
The complete EEND system is re-finetuned using the powerset loss with the pruned WavLM restored in place. The same learning rates as in Stage~1 are used. Early stopping is applied with a patience of 5~epochs, typically converging within 10~epochs.

To obtain global diarization results, we extract local speaker embeddings using a ResNet34-LM model trained with the WeSpeaker toolkit~\cite{wang2024advancing} on the VoxCeleb2 dataset~\cite{chung2018voxceleb2}. For models trained on the compound dataset comprising AMI, AISHELL-4, and AliMeeting, agglomerative hierarchical clustering is applied. For models trained on the larger compound dataset described in Section~\ref{subsec: datasets}, VBx clustering~\cite{landini2022bayesian} is used instead. Detailed configurations are available in our repository.\footnote{\url{https://github.com/BUTSpeechFIT/DiariZen}}

% In this case, the initial AHC threshold is set to 0.6, the acoustic scaling factor $\rm F_A$ is 0.07, and the speaker regularization coefficient $\rm F_B$ is 0.8.

For evaluation, we report DER without collar on all datasets except CHiME-6.  
For CHiME-6, we additionally follow the official evaluation protocol and report DER with a 0.25\,s collar. We further report the macro-averaged DER to reflect overall performance across all datasets.

% Our source code is publicly available\footnote{\url{https://github.com/BUTSpeechFIT/DiariZen}}.

% \begin{table}[tbp]
% \vspace{-0.1cm}
%   \caption{Ablation analysis of knowledge distillation. Sparsity is 90\%. WavLM base+ is used. ``Trans'' means the linear transformation from student to teacher.}
%   \label{tab:ablation_distill}
%  \setlength{\tabcolsep}{0.4mm}
%   \centering
%   \begin{tabular}{l| c | c c c | c}
%     \hline
%         \multirow{2}{*}{Layer Index} & \multirow{2}{*}{Trans.} & \multicolumn{3}{c|}{DER (\%)} & \multirow{2}{*}{Macro} \\
%         &  & AMI & AISHELL-4 & AliMeeting &  \\
%     \hline
%     0,12 & - &  &  &  &  \\
%     0,4,8,12 & - &  &  &  &  \\
%     \hline
%     \multirow{2}{*}{full layers} & - &  &  &  &  \\
%         & \checkmark &  &  &  &  \\  
%     \hline
%   \end{tabular}
%   \vspace{-0.1cm}
% \end{table}

% \begin{table}[ht]
% \centering
% \caption{Performance when training models with different data ratios. \\
% The pruning objective is the number of parameters and sparsity is set to 80\%. WavLM base+ is used.}
% \begin{tabular}{ccccc c}
% \toprule
% \raisebox{4ex}{Ratio} & \raisebox{4ex}{Hours} & \rotatebox{90}{AMI} & \rotatebox{90}{AISHELL-4} & \rotatebox{90}{AliMeeting} & \rotatebox{90}{Macro} \\
% \midrule
% 100\% & 288.3 & 15.8 & 12.1 & 17.9 & 15.3 \\
% 75\%  & 216.2 & 16.0 & 12.0 & 17.8 & 15.3 \\
% 50\%  & 144.1 & 16.2 & 12.5 & 18.5 & 15.7 \\
% 25\%  &  72.1 & 16.6 & 12.1 & 17.8 & 15.5 \\
% 5\%   &  14.1 & 19.1 & 14.2 & 21.8 & 18.4 \\
% \bottomrule
% \end{tabular}
% \label{tab:data-ratio}
% \end{table}

\begin{table*}[htbp]
\vspace{-0.1cm}
  \caption{Performance comparison under different sparsity setups. The pruning objective is the number of parameters and sparsity is 80\%.}
  \label{tab:cnn_trans_separate}
 \setlength{\tabcolsep}{1.4mm}
  \centering
  \begin{tabular}{l| c | c c c | c c c | c c | c c c | c}
    \hline
        \multirow{2}{*}{System} & \multirow{2}{*}{Sparsity} & \multicolumn{3}{c|}{Params} & \multicolumn{3}{c|}{MACs} & \multicolumn{2}{c|}{Speedup$\nearrow$} & \multicolumn{3}{c|}{DER (\%)} & \multirow{2}{*}{Macro} \\
        & & All & CNNs & Trans & All & CNNs & Trans & GPU & CPU & AMI & AISHELL-4 & AliMeeting &  \\
    \hline
    \multirow{3}{*}{WavLM Base+} & - & 94.4M & 4.2M & 90.2M & 6.9G & 2.4G & 4.5G & - & - & 15.8 & 12.1 & 17.9 & 15.3 \\
    & overall & 18.8M & 0.6M & 18.2M & 1.1G & 0.2G & 0.9G & 4.0$\times$ & 4.5$\times$ & 15.8 & 12.1 & 17.9 & 15.3 \\
     & separate & 18.9M & 0.9M & 18.1M & 1.2G & 0.3G & 0.9G & 3.9$\times$ & 4.4$\times$ & 15.7 & 11.8 & 17.9 & 15.1 \\
     \hline
    \multirow{3}{*}{WavLM Large} & - & 315.6M & 4.2M & 311.4M & 17.8G & 2.4G & 15.4G & - & - & 15.1 & 11.3 & 15.8 & 14.1 \\
    & overall & 63.3M & 1.4M & 61.9M & 3.8G & 0.7G & 3.1G & 2.6$\times$ & 3.1$\times$ & 15.1 & 11.3 & 15.8 & 14.1 \\
     & separate & 64.5M & 0.8M & 63.7M & 3.6G & 0.5G & 3.1G & 2.6$\times$ & 3.8$\times$ & 15.2 & 11.2 & 17.5 & 14.6 \\
    \hline
  \end{tabular}
  \vspace{-0.1cm}
\end{table*}

\section{Results and Discussions}
\subsection{Overall performance}
\label{subsec: overall}
% \begin{itemize}
%     \item Table \ref{tab:overall1}, an overview, just shows the results when sparsity is 0/80/90\%. Our method works well for both WavLM Base+ and WavLM Large. Our method achieves comparable performance to the unpruned model when removing 80\% redundant parameters. 
%     However, the speedup of WavLM Large is not as significant as that of WavLM Base+.
    
%     \item Figure \ref{fig:cnn_trans}, absolute inference time of CNN/Transformer after pruning; clearly speedup; different patterns for WavLM Base+ and WavLM Large. Params of CNN and Transformer; CNNs seem to be a bottleneck of the large model.
% \end{itemize}

% We show the pruning performance of WavLM models when sparsities are set to 0/80/90\% in Table \ref{tab:overall1}. 

Table~\ref{tab:overall1} presents the overall performance of the WavLM-based diarization system under different sparsity levels. Both the number of parameters and the number of MACs are considered as pruning objectives. For clearer comparison, we also report the post-pruning characteristics of each WavLM model, including the number of parameters, MACs, and inference speedup relative to the unpruned model. Further discussion is provided in Section~\ref{subsec: cnn_trans_inference_time}. To ensure a robust evaluation of inference acceleration, each setting is executed five times, and the average inference speedup is reported.

For reference, results using traditional filterbank (Fbank) features as input to the Conformer model are also provided. As shown, WavLM-based models consistently outperform those using Fbank features, with the Large model achieving better results than the Base+ one in all settings. Both pruned WavLM Base+ and Large models retain performance comparable to their unpruned counterparts even at 80\% sparsity, and this trend remains consistent across different pruning objectives. At 90\% sparsity, a higher inference speedup can be achieved, albeit typically at the cost of some performance degradation.
\reviewertwo{
From a computational perspective, the Fbank- and WavLM-based systems share the same diarization backend and differ only in the front-end feature extraction, i.e., Fbank versus WavLM.
For a 30-minute input recording processed on a single NVIDIA RTX A5000 GPU, the unpruned WavLM Base+ model requires 27.3~s for feature extraction. After applying structured pruning with 80\% sparsity, the processing time is reduced to 7.2~s, substantially narrowing the computational gap relative to the Fbank system.
}

In terms of diarization performance, there is no significant difference between using the number of parameters or MACs as the pruning criterion. However, the pruning behavior is strongly influenced by the selected objective, with each criterion more effectively constraining its respective target metric. For example, when pruning is guided by MACs, the resulting models typically achieve the desired MAC count, but the number of parameters often becomes difficult to control. Interestingly, pruning based on parameter count generally tends to yield MACs comparable to those achieved by pruning directly for MACs, and in the case of WavLM Base+, sometimes even lower. In contrast, pruning for MACs frequently results in models with substantially higher parameter counts. These findings suggest that using the number of parameters as the pruning objective may offer a more favorable trade-off between model efficiency and performance. We therefore adopt it as the default pruning criterion in subsequent experiments.

Besides, our experiments reveal distinct pruning patterns between WavLM Base+ and WavLM Large. When pruning by parameter count at the same sparsity level, the pruned Base+ model achieves significantly higher inference acceleration than the pruned Large model (e.g. when using a single GPU, 4.0$\times$ vs. 2.6$\times$; 5.9$\times$ vs. 3.4$\times$). However, this discrepancy vanishes when MACs are used for pruning: in such cases, the pruned Base+ models often become slower, while the Large models benefit from considerable speedups.  Additionally, although MACs are generally correlated with computational complexity, they are not always reliable indicators of actual inference speed. For instance, when pruning WavLM Large using either objective, the resulting models may exhibit similar MACs but differ substantially in real-world acceleration.

\subsection{Inference time for CNN and Trans. layers}
\label{subsec: cnn_trans_inference_time}
\begin{figure}[tbp]
% \vspace{-0.5cm}
  \centering
  \includegraphics[width=8cm]{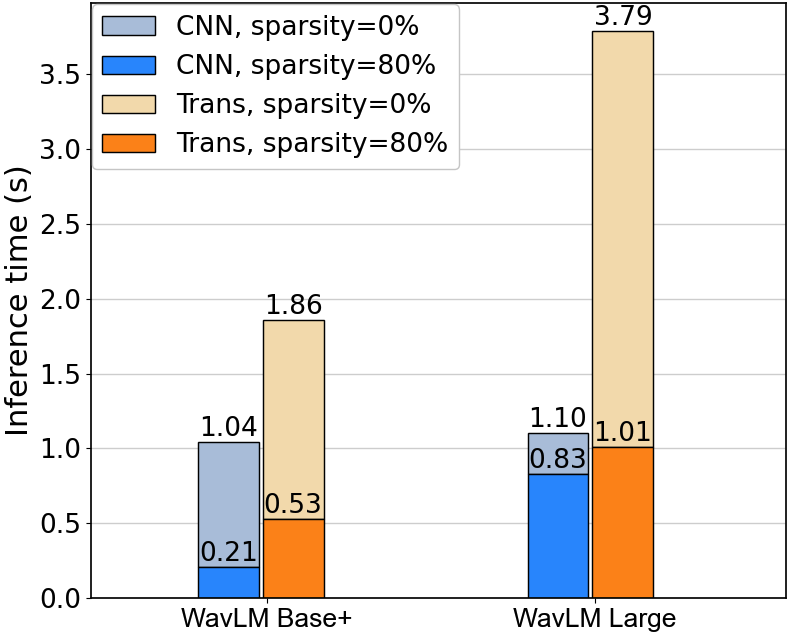}
  \caption{Absolute inference time comparison between CNN and Transformer layers, with the number of parameters as the pruning objective. All reported numbers are averaged over five runs.}
  \label{fig:cnn_trans}
  \vspace{-0.3cm}
\end{figure}

As illustrated, WavLM \cite{chen2022wavlm} consists of several CNN layers followed by Transformer (Trans.) layers.
In Figure~\ref{fig:cnn_trans}, we compare the absolute inference times of the CNN and Transformer layers separately for both unpruned and pruned (80\% sparsity) WavLM Base+ and Large models, with parameter count as the pruning objective. 
Each input sample corresponds to an 8-second local window in the EEND-VC framework, and the batch size is adjusted to fully utilize GPU resources. Since the Base+ model has fewer parameters than the Large model, it supports a larger batch size. Specifically, we set batch size to 230 for the Base+ and 150 for the Large models.
As we can see, although CNNs contain a small number of parameters (4.2 million, 4.4\% of all parameters for WavLM Base+ and 1.3\% of all parameters for WavLM Large; see Table \ref{tab:cnn_trans_separate}), they are quite computationally intensive and slow during inference, accounting for 35.9\% and 22.5\% of the total inference time for the Base+ and  Large models, respectively. 
After pruning,
it is clear that both models achieve notable and similar acceleration in the Transformer layers. 
% However, the Base+ model eliminates more CNN parameters compared to the Large model, leading to a higher speedup (4.0$\times$ vs. 2.6$\times$).
\reviewertwo{However, as shown by the overall
pruning results reported in Table~\ref{tab:cnn_trans_separate}, the Base+ model removes more CNN
parameters than the Large model, which likely contributes to a higher overall speedup
(4.0$\times$ vs.\ 2.6$\times$).}
% Since pruning is performed in a fully data-driven manner, one possible explanation for this difference is that the CNN representations in WavLM Large are denser and more informative than in the Base+ model, making it a lower priority for pruning.

\subsection{Separate sparsity constraints for CNN and Trans. layers}
\label{subsec: cnn_trans_separate_pruning}

As previously shown, CNNs have a significant impact on inference speed. We also observe that WavLM Base+ and Large exhibit different pruning behaviors for CNN and Transformer layers. Given these differences, it is natural to ask whether explicitly encouraging the WavLM Large model to prune more CNN layers can further accelerate inference.

To explore this, we impose separate sparsity constraints on the CNN and Transformer components by modifying the Lagrangian terms in the final loss as
\begin{align}
    \sum_{m \in \{cnn, trans\}} 
        \lambda_1 \big(\Vert \bm{\theta}_m \Vert_0 - t\big)
        + \lambda_2 \big(\Vert \bm{\theta}_m \Vert_0 - t\big)^2 . \notag
\end{align}
The results are reported in Table~\ref{tab:cnn_trans_separate}, where sparsity is computed for the overall model as well as separately for the CNN and Transformer layers.

Compared with pruning based on overall parameter sparsity, the separate constraint removes parameters from CNN and Transformer layers more evenly but provides no clear advantage. For WavLM Base+, it even results in a higher MAC count. For WavLM Large, although fewer CNN parameters and MACs are retained, the inference speed on a single GPU remains unchanged and the overall performance degrades. 
\reviewertwo{This behavior can be attributed to system-level factors: for the Large model,
inference time is dominated by the Transformer layers, and reductions in CNN
parameters and MACs alone may not translate into additional GPU wall-clock
speedups. At the same time, enforcing separate sparsity constraints across CNN and Transformer components limits the ability to optimally allocate sparsity in a global manner, which may introduce unnecessary
constraints on representation learning and lead to degraded diarization
performance.}
These results indicate that the original overall pruning is more effective at preserving model performance. We therefore adopt the overall pruning strategy in subsequent experiments.

\subsection{Analysis under different sparsity levels}
\label{subsec: analysis_different_sparsity}

As shown in Table \ref{tab:overall1}, leveraging a pre-trained large model consistently proves advantageous. For example, at 80\% sparsity, the pruned WavLM Large contains fewer parameters than the unpruned WavLM Base+ (63.3M vs. 94.4M) while delivering significantly better performance.
Note that both WavLM Base+ and WavLM Large are pre-trained using the same corpus, which 
raises a natural question: if our goal is to obtain a small  model, is it always beneficial to use the largest pre-trained model and then prune it to the desired size?

\begin{figure}[tbp]
% \vspace{-0.1cm}   
  \centering
  \includegraphics[width=8.2cm]{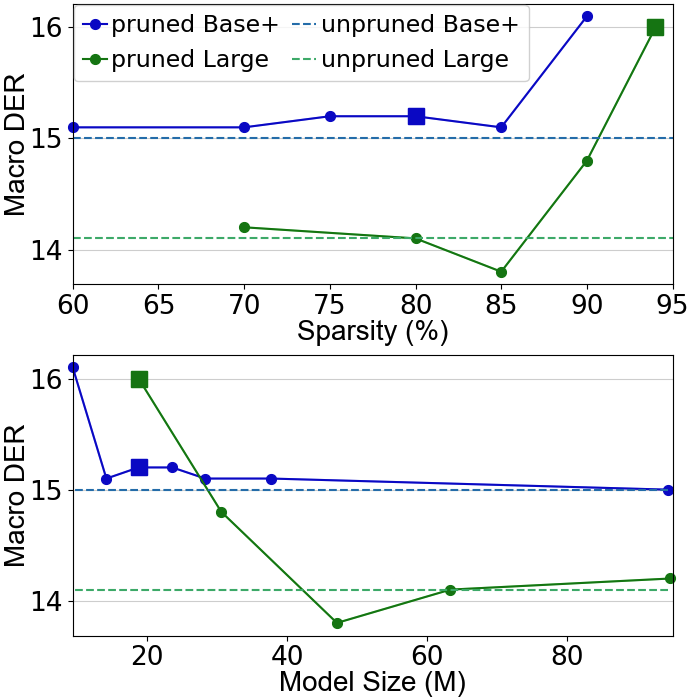}
  \caption{
  Performance comparisons between WavLM Base+ and WavLM Large under different pruning setups. The blue and green rectangles indicate that the pruned Base+ and Large models have a similar number of parameters. }
  \label{fig:sparsity_params_base_large}
  \vspace{-0.1cm}   
\end{figure}

To address the above question, we evaluate the pruning performance of WavLM Base+ and WavLM Large under various sparsity levels, as illustrated in Figure~\ref{fig:sparsity_params_base_large}. As indicated, for the same pruning sparsity, the pruned WavLM Large consistently outperforms its Base+ counterpart. Furthermore, both the Large and Base+ models can retain similar performance to the unpruned models even after removing 85\% of their parameters. However, at 94\% sparsity, where the pruned WavLM Large (green rectangle) has the same number of parameters (18.8M) as the Base+ model (blue rectangle) pruned at 80\%, its performance degrades significantly.
Our results suggest that excessive pruning can completely undermine a model's performance. Thus, in scenarios requiring a compact and efficient model, pruning a smaller model may be more reasonable than pruning a larger one.

% \begin{figure}[tbp]
% % \vspace{-0.5cm}
%   \centering
%   \includegraphics[width=8cm]{sparisty_pruning_ratio_cnn_trans.png}
%   \caption{Pruning ratios of CNNs and Transformer layers under varying sparsity levels. The pruning objective is the number of parameters.}
%   \label{fig:pruning_ratio_cnn_trans}
%   \vspace{-0.3cm}
% \end{figure}

% \vspace{-0.5mm}

\subsection{Ablation analysis}
\label{subsec: ablation_analysis}

\subsubsection{Progressive pruning from already pruned models}
Our results in Table~\ref{tab:overall1} and Figure~\ref{fig:sparsity_params_base_large} demonstrate that our method can achieve performance comparable to the unpruned model even at high sparsity levels. However, in a new application scenario, it is often difficult to determine how many parameters can be safely removed without causing significant performance degradation. 
One straightforward approach is to run multiple pruning experiments at different sparsity levels and select the best outcome. However, it is widely recognized that knowledge distillation-based training is resource-intensive, meaning that obtaining a well-performing pruned model for a new task can be costly.
A more flexible alternative is to first prune the model to a relatively conservative sparsity level and then incrementally increase the sparsity based on the already pruned model. 
In this framework, the conservatively pruned model serves as the teacher for subsequent pruning stages.
Since both the teacher and student models have already undergone pruning, their parameter counts and MACs are significantly reduced. As a result, subsequent training toward higher sparsity becomes faster and more computationally efficient.

To explore this idea, Table~\ref{tab:pruning_setup} presents an experiment where we obtain 80\% and 90\% sparsity models either directly from the unpruned model (0\% sparsity) or from a model that has  been pruned to 60\% sparsity. As shown, the models obtained by further pruning from already pruned models exhibit nearly identical performance to those pruned from scratch. These results highlight the potential of progressive pruning as a flexible and practical solution for real-world deployment.

\begin{table}[tbp]
\vspace{-0.1cm}
  \caption{Performance under different pruning setups. The pruning objective is the number of parameters. WavLM base+ is used.}
  \label{tab:pruning_setup}
 \setlength{\tabcolsep}{0.8mm}
  \centering
  \begin{tabular}{l| c | c | c c c | c}
    \hline
        \multirow{2}{*}{Sparsity} & \multirow{2}{*}{Params} & \multirow{2}{*}{MACs} & \multicolumn{3}{c|}{DER (\%)} & \multirow{2}{*}{Macro} \\
        & &  & AMI & AISHELL-4 & AliMeeting &  \\
    \hline
    0 --> 60\% & 37.7M & 2.3G & 15.8 & 11.7 & 17.7 & 15.1 \\
    0 --> 80\% & 18.8M & 1.1G & 15.8 & 12.1 & 17.9 & 15.3 \\
    0 --> 90\% & 9.4M & 0.6G & 16.9 & 11.8 & 18.5 & 15.7 \\
    0 --> 60 --> 80\% & 18.9M & 1.2G & 15.6 & 12.3 & 18.1 & 15.3 \\
    0 --> 60 --> 90\% & 9.4M & 0.6G & 16.6 & 12.0 & 18.6 & 15.7 \\
    \hline
  \end{tabular}
  \vspace{-0.3cm}
\end{table}

\begin{table}[!htbp]
\vspace{-0.1cm}
  \caption{Effects of extended training at 90\% sparsity. The pruning objective is the number of parameters. WavLM base+ is used.}
  \label{tab:training_epochs}
  \centering
  \begin{tabular}{l | c c c | c}
    \hline
        \multirow{2}{*}{Training Epochs} & \multicolumn{3}{c|}{DER (\%)} & \multirow{2}{*}{Macro} \\
      & AMI & AISHELL-4 & AliMeeting &  \\
    \hline
    warmup=5, total=30 & 16.9 & 11.8 & 18.5 & 15.7 \\
    warmup=10, total=35 & 16.6 & 12.5 & 18.6 & 15.9 \\
    warmup=15, total=40 & 16.6 & 12.2 & 18.4 & 15.8 \\
    \hline
  \end{tabular}
  \vspace{-0.3cm}
\end{table}

\begin{table}[htbp]
  \caption{Comparison of pruning strategies for WavLM Base+ at 80\% sparsity. 
  \textit{preFT} indicates whether WavLM was fine-tuned on the diarization task before pruning.}
  \label{tab:ablation}
  \setlength{\tabcolsep}{1.4mm}
  \centering
  \begin{tabular}{l|c|ccc|c}
    \hline
    \multirow{2}{*}{Pruning strategy} & \multirow{2}{*}{preFT} & \multicolumn{3}{c|}{DER (\%)} & \multirow{2}{*}{Macro} \\
     &  & AMI & AISHELL-4 & AliMeeting &  \\ \hline

    Unpruned baseline & -- & 15.6 & 11.8 & 17.7 & 15.0 \\ \hline

    Prune w/  & --         & 16.7 & 12.0 & 18.3 & 15.7 \\
    diarization loss & \checkmark & 17.0 & 12.1 & 19.7 & 16.3 \\ \hline

    \multirow{2}{*}{Prune w/ distillation}      & --         & 16.5 & 11.9 & 19.5 & 16.0 \\
                                              & \checkmark & 15.7 & 12.1 & 19.1 & 15.6 \\
    \quad + post-distillation & \checkmark & 15.8 & 12.1 & 17.9 & 15.3 \\ \hline
  \end{tabular}
  \vspace{-0.1cm}
\end{table}

\subsubsection{Extended training at 90\% sparsity}
As discussed earlier, system performance typically degrades at high sparsity levels such as 90\%. Even when the warm-up strategy is employed, which linearly increases the target sparsity over a certain number of epochs, the target sparsity at each warm-up step may still be too high to preserve the original unpruned performance. In addition, higher sparsity levels may require longer training schedules to allow the model to better adapt.

To investigate these assumptions, we evaluate different training setups for pruning WavLM Base+ at 90\% sparsity, as shown in Table~\ref{tab:training_epochs}. As observed, although finer-grained sparsity warm-up strategies and extended training epochs are applied, the performance remains largely unchanged. These findings suggest that simply increasing training time or smoothing the sparsity schedule does not effectively mitigate the performance degradation at high sparsity levels.

\subsubsection{Comparison of pruning alternatives}
\label{subsub: pruning_alternatives}
Table~\ref{tab:ablation} compares alternative pruning strategies for WavLM Base+ at a fixed 80\% sparsity. The first row reports the unpruned baseline. The last row corresponds to the pruning strategy used in our main results (same as the line \textit{WavLM Base+ 80\% Params} in Table~\ref{tab:overall1}), i.e., the three-stage pipeline: Stage~1 trains the entire EEND system on the diarization task to adapt WavLM; Stage~2 prunes WavLM under a distillation objective and then continues distillation without further pruning as described in Sec.~\ref{subsec: configurations}; and Stage~3 re-finetunes the EEND system after pruning. The same Stage~3 re-finetuning is applied to all alternatives in Table~\ref{tab:ablation}.
The  \textit{preFT} indicates whether Stage~1 (diarization pre-fine-tuning before pruning) was performed.

As a Stage~2 alternative, rows~2--3 apply pruning directly under the diarization (powerset) loss, meaning that the entire EEND model is optimized while enforcing sparsity via $L_0$ regularization. This task-loss-based approach, commonly used in pruning for other tasks~\cite{peng2023structured}, produces the largest degradation relative to the unpruned model, regardless of whether Stage~1 pre-fine-tuning is used.

Rows~4--5 prune WavLM under a distillation loss instead of the diarization loss. When pruning starts from an unadapted WavLM (\textit{preFT=--}), performance remains clearly below the baseline. Initializing both teacher and student from a diarization-adapted WavLM (\textit{preFT=\checkmark}) narrows this gap, although the result on AliMeeting remains below the baseline. 

Finally, in the last row, once pruning has been completed and the target sparsity is in effect, distillation is continued for about 15~epochs with pruning frozen (see Sec.~\ref{subsec: configurations}). This post-pruning distillation step recovers most of the remaining gap and yields the strongest pruned model, which is why it is adopted as our default approach.

\begin{comment}
\subsubsection{Comparison of pruning alternatives}
\label{subsub: pruning_alternatives}
% \vspace{-1mm}

As shown in Table~\ref{tab:ablation}, we compare several approaches for applying structured pruning to  WavLM Base+ at an 80\% sparsity level.
The unpruned results are also provided in the first row as a reference.
Rows 2-6 present different pruning strategies, where each pruned WavLM model is further fine-tuned on the diarization task. 
Specifically, Rows 2 and 3 (diar+pruning) correspond to directly fine-tuning the WavLM models following the diarization system shown in Figure~\ref{fig:framework}(a), using a combination of powerset loss \cite{plaquet2023powerset} and $L_0$ regularization.
Rows 4 and 5 (distill+pruning) show results where the WavLM models are pruned with both distillation and pruning losses, which constitutes our default pruning strategy.  The final row reports results after additional distillation (up to 20 epochs) from the original teacher.
As we can see, directly applying the diarization loss and pruning loss to WavLM leads to suboptimal performance.
When incorporating distillation loss during pruning, it is crucial to initialize both the teacher and student models with WavLM fine-tuned for the diarization task (preFT) to preserve unpruned performance on AMI. However, the performance on AliMeeting still lags significantly behind that of the unpruned model.  
This gap can be bridged by further distilling the pruned model to the original teacher.
\end{comment}

\begin{figure}[tbp]
% \vspace{-0.5cm}
  \centering
  \includegraphics[width=8cm]{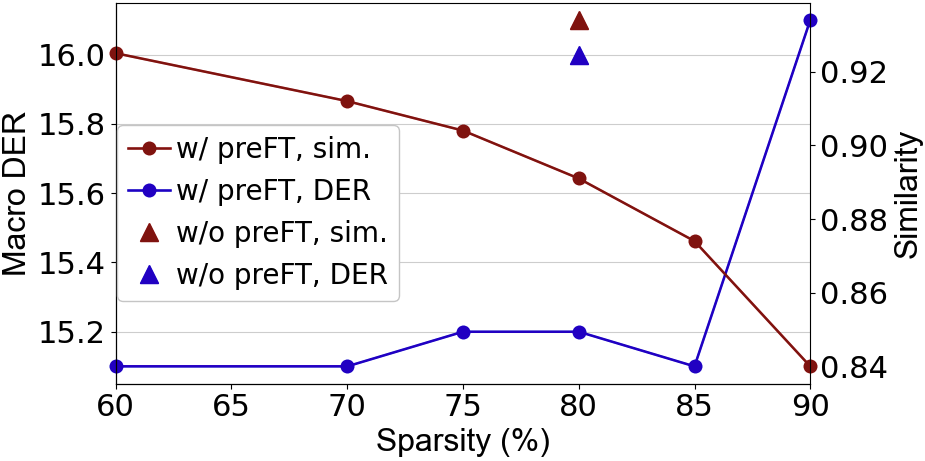}
  \caption{Macro DER and cosine similarity between the layer outputs of the pruned model and its teacher for WavLM Base+ under different pruning setups. The triangle points show the performance for 80\% sparsity when distilling knowledge from the original (not fine-tuned) WavLM.}
  \label{fig:cos_sim}
  % \vspace{-0.5cm}
\end{figure}

\subsubsection{Teacher–student representation similarity}
Figure~\ref{fig:cos_sim} shows the cosine similarity between the layer outputs of the pruned model and its teacher, along with the corresponding DER at different sparsity levels. The cosine similarity is averaged over
Transformer layers \{0, 4, 8, 12\} and computed on the full development set. When the teacher is the Stage~1 (pre-fine-tuned) WavLM Base+, the similarity decreases as sparsity increases but remains relatively high even at 90\% sparsity. Notably, a lower similarity does not necessarily imply worse DER when sparsity is below 85\%. Beyond 85\% sparsity, both similarity and DER degrade in a coupled fashion.

In contrast, when the original (unadapted) WavLM Base+ is used as the teacher---shown by the triangle markers---the pruned model still achieves high cosine similarity but performs much worse in terms of DER. This demonstrates that matching the teacher's internal representations is not sufficient unless the teacher itself has been pre-fine-tuned to the diarization task beforehand. In other words, WavLM pre-fine-tuning (Stage~1) is crucial for effective pruning.

\subsubsection{Effects of data quantity}
\label{subsubsec: data_quantity}

% In our previous work \cite{han2025leveraging}, we demonstrated that a WavLM-based diarization model can maintain competitive performance even when trained on only half of the available data. In the current pruning setup, the default strategy utilizes the full training set for model pruning. However, obtaining high-quality annotated data continues to pose a major challenge in many real-world applications. Moreover, training models with a knowledge distillation objective tends to be computationally intensive. This motivates an investigation into how varying the amount of training data influences the diarization performance of pruned models. One promising direction for reducing overall training cost is to limit the amount of data used during the pruning stage.

In our previous work \cite{han2025leveraging}, we demonstrated that a WavLM-based diarization model can maintain competitive performance even when trained on only half of the available data. In the current pruning setup, the default strategy utilizes the full training set for model pruning. \reviewertwo{However, high-quality annotated data remain scarce in many real-world scenarios, and training with knowledge distillation is computationally intensive.} This motivates an investigation into how varying the amount of training data influences the diarization performance of pruned models. One promising direction for reducing overall training cost is to limit the amount of data used during the pruning stage.

\begin{figure}[tbp]
% \vspace{-0.5cm}
  \centering
  \includegraphics[width=8.8cm]{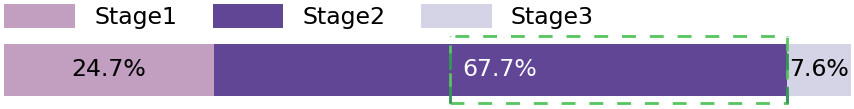}
  \caption{Relative time consumption of WavLM Base+ across different stages of the pipeline when using the full dataset. The dashed green box  represents the time saved when using half of the dataset for model pruning.}
  \label{fig:time_consumption}
  \vspace{-0.3cm}
\end{figure}

% To this end, we analyze the relative time consumption across different training stages, as shown in Figure~\ref{fig:time_consumption}. 
% The pruning objective is defined by the number of parameters, with the target sparsity set to 80\%. 
% Stage~1 adapts the WavLM component to the diarization task by training the complete EEND model with the powerset loss; Stage~2 prunes only the WavLM component using a distillation objective and then continues distillation without further pruning (see Sec.~\ref{subsec: configurations}); and Stage~3 re-finetunes the complete EEND model with the pruned WavLM to recover any remaining performance loss. 
\reviewertwo{To this end, we analyze the relative time cost across different training stages in Figure~\ref{fig:time_consumption}. 
As observed, most of the training cost is incurred during Stage~2 (67.7\% of total), while Stage~3 accounts for only 7.6\%, motivating
the use of reduced training data during pruning and reserving more data for final fine-tuning. 
Based on this observation, we study the impact of training data size on pruning
and subsequent fine-tuning by sampling 50\%, 25\%, and 5\% subsets from AMI,
AISHELL-4, and AliMeeting. For each setting, we train WavLM Base+ models with
80\% sparsity, while the teacher model is always trained on the full dataset.
Results are reported in Table~\ref{tab:data_ratio_prune_ft}.}
% which also reports the number of epochs required for convergence in both the pruning stage (including distillation-guided pruning and further distillation) and the subsequent fine-tuning.

% \begin{table}[tbp]
% \centering
% \begin{threeparttable}
%   \caption{Performance when training models with different data ratios at the pruning and further fine-tuning (furFT) stages.}
%   \label{tab:data_ratio_prune_ft}
%  \setlength{\tabcolsep}{1.0mm}
%   \begin{tabular}{l c| c c | c c c | c}
%     \hline
%         \multicolumn{2}{c|}{Pruning} & \multicolumn{2}{c|}{FurFT} & \multicolumn{3}{c|}{DER (\%)} & \multirow{2}{*}{Macro} \\
%      Ratio & Epoch & Ratio & Epoch & AMI & AISHELL-4 & AliMeeting &  \\
%     \hline
%     100\% & 30 + 6 & 100\% & 8 & 15.8 & 12.1 & 17.9 & 15.3 \\
%     \hline
%     \multirow{3}{*}{50\%} & \multirow{3}{*}{30 + 7} & 50\% & 3 & 16.2 & 12.5 & 18.5 & 15.7 \\
%         & & 50\%\tnote{*} & 8 & 17.0 & 12.0 & 18.8 & 15.9 \\
%       & &  100\% & 8 & 15.8 & 12.0 & 17.6 & 15.1 \\
%     \hline 
%     \multirow{2}{*}{25\%} & \multirow{2}{*}{30 + 20} & 25\% & 7 & 16.6 & 12.1 & 17.8 & 15.5 \\
%      &  & 100\% & 8 & 15.9 & 12.1 & 17.9 & 15.3 \\
%     \hline 
%     \multirow{2}{*}{5\%} & \multirow{2}{*}{6 + 20} & 5\% & 3 & 19.1 & 14.2 & 21.8 & 18.4 \\
%      &  & 100\% & 20 & 16.3 & 12.0 & 19.3 & 15.9 \\
%     \hline 
%   \end{tabular}
%   \begin{tablenotes}
%     \footnotesize
%     \item[*] Only the remaining 50\% of the data is used.
%     \end{tablenotes}
%   \end{threeparttable}
%   \vspace{-0.2cm}
% \end{table}

As shown, our method achieves comparable performance even when both pruning and fine-tuning are performed on reduced subsets. Although performance degradation is observed in some cases, it is largely compensated when the full dataset is used for the final fine-tuning stage. We further examined the case in which fine-tuning uses only the portion of the data not seen during pruning (e.g., for the 50\% pruning condition, the remaining 50\% is used for fine-tuning). This configuration yields slightly worse performance compared with using the same data as pruning. Fine-tuning on the entire dataset consistently leads to the best results. Notably, pruning with only 5\% of the data still yields reasonable performance.

\begin{table}[tbp]
\centering
\begin{threeparttable}
  \caption{Performance when training models with different data ratios at the pruning and further fine-tuning (furFT) stages.}
  \label{tab:data_ratio_prune_ft}
 \setlength{\tabcolsep}{2mm}
  \begin{tabular}{l| c | c c c | c}
    \hline
        \multirow{2}{*}{Pruning} & \multirow{2}{*}{FurFT} & \multicolumn{3}{c|}{DER (\%)} & \multirow{2}{*}{Macro} \\
      & & AMI & AISHELL-4 & AliMeeting &  \\
    \hline
    100\% & 100\% & 15.8 & 12.1 & 17.9 & 15.3 \\
    \hline
    \multirow{3}{*}{50\%} & 50\% & 16.2 & 12.5 & 18.5 & 15.7 \\
        & 50\%\tnote{*} & 17.0 & 12.0 & 18.8 & 15.9 \\
        & 100\% & 15.8 & 12.0 & 17.6 & 15.1 \\
    \hline 
    \multirow{2}{*}{25\%} & 25\%  & 16.6 & 12.1 & 17.8 & 15.5 \\
        & 100\% & 15.9 & 12.1 & 17.9 & 15.3 \\
    \hline 
    \multirow{2}{*}{5\%} & 5\% & 19.1 & 14.2 & 21.8 & 18.4 \\
        & 100\% & 16.3 & 12.0 & 19.3 & 15.9 \\
    \hline 
  \end{tabular}
  \begin{tablenotes}
    \footnotesize
    \item[*] Only the remaining 50\% of the data is used.
    \end{tablenotes}
  \end{threeparttable}
  \vspace{-0.2cm}
\end{table}

\begin{table*}[t]
\centering
\begin{threeparttable}
\caption{Performance across different datasets with varying local EEND window lengths during training and inference. No domain adaptation; all system uses the same clustering hyper-parameters. The evaluation collar is set to 0 seconds for all datasets. The best results are shown in bold. }
\label{tab:compound_results}
\setlength{\tabcolsep}{1.4mm}
\begin{tabular}{l |c | cc |cccccccc | c}
\hline
\multirow{2}{*}{System} & \multirow{2}{*}{Sparsity} & \multicolumn{2}{c|}{Window} & \multicolumn{8}{c|}{DER (\%)} & \multirow{2}{*}{Macro} \\
& & Train & Infer & AMI & AIS-4 & AliM & NSF & MSD & DH3 & RAMC & VoxC &  \\
\hline
\multirow{6}{*}{WavLM Base+} & 0\% & 8s & 8s & 16.3 & 10.9 & 15.8 & 19.3 & 17.8 & 16.5 & 10.9 & 9.6 & 14.6 \\
& 0\% &  8s & 12s & 16.1 & 10.5 & 16.0 & 19.7 & 17.5 & 16.4 & 11.2 & 9.5 & 14.6 \\
& 80\% &  8s & 8s & 16.7 & 11.2 & 16.2 & 20.7 & 18.0 & 16.9 & 11.5 & 10.1 & 15.2 \\
& 80\% &  8s & 12s & 16.2 & 10.9 & 16.3 & 21.4 & 18.0 & 16.8 & 11.7 & 10.0 & 15.2 \\
& 80\% &  12s & 12s & 15.9 & 11.1 & 15.8 & 19.8 & 17.6 & 16.1 & 11.3 & 9.8 & 14.7 \\
& 80\% &  16s & 16s & 15.8 & 10.7 & 14.1 & 20.3 & 17.4 & 15.9 & 11.4 & 9.7 & 14.4 \\
\hline
\multirow{5}{*}{WavLM Large} & 0\% & 8s & 8s & 15.7 & 10.6 & 15.0 & 18.2 & 16.6 & 15.4 & 11.1 & 9.5 & 14.0 \\
& 80\% & 8s & 8s & 15.1 & 10.3 & 15.1 & 18.3 & 16.8 & 15.2 & 11.1 & 9.3 & 13.9 \\
& 80\% &  12s & 12s & 14.9 & 10.2 & 13.9 & 18.0 & 16.2 & 14.9 & 11.0 & 9.2 & 13.5 \\
& 80\% &  16s & 16s & 14.0 & \textbf{9.8} & 12.5 & 17.9 & \textbf{15.6} &  \textbf{14.5} & 11.0 & 9.2 & 13.1 \\
\qquad w/ 4spk overlap powerset & 80\% &  16s & 16s & \textbf{13.9} & 10.1 & \textbf{10.8} & \textbf{16.7} & 15.8 & \textbf{14.5} & 11.0 & \textbf{9.1} & \textbf{12.7} \\
% & 80\% &  16s & 16s & 15.7 & 10.6 & 15.9 & 23.5 & 21.7 & 19.9 & 12.2 & 10.1 & 16.2 \\
\hline
State-of-the-art by submission\tnote{*} & - & - & - & 15.2 \cite{broughton2025pushing} & 10.2 \cite{plaquet2025dissecting} & 11.4 \cite{broughton2025pushing} & 19.7 \cite{niu2024dcf} & 17.3 \cite{plaquet2025dissecting} & 14.5 \cite{broughton2025pushing} & \textbf{10.7} \cite{plaquet2025dissecting} & 9.3\tnote{\dag} \cite{plaquet2024mambabasedsegmentationmodelspeaker} & - \\
\hline
\end{tabular}
  \begin{tablenotes}
    \footnotesize
    \item[*] Potentially over-optimistic, as state-of-the-art results are typically achieved through fine-tuning on each specific dataset.
    \item [\dag] For consistency, we report all state-of-the-art results without a collar. Many prior works use a 0.25-s collar, so their numbers are not directly comparable.
    \end{tablenotes}
\label{tab:system-performance}
  \vspace{-0.2cm}
\end{threeparttable}
\end{table*}

In summary, using 50\% or 25\% of the data for pruning, followed by fine-tuning on the full dataset, provides high training efficiency with no observed performance degradation. As highlighted by the green dashed box in Figure~\ref{fig:time_consumption}, the 50\% setting significantly reduces overall training time. These results demonstrate that the proposed framework can be effectively applied in resource-constrained scenarios, reducing reliance on large-scale labeled data while maintaining strong performance.

\subsubsection{Visualization of Layer-wise Pruning Patterns}
\begin{figure}[tbp]
% \vspace{-0.1cm}
  \centering
 \includegraphics[width=8cm]{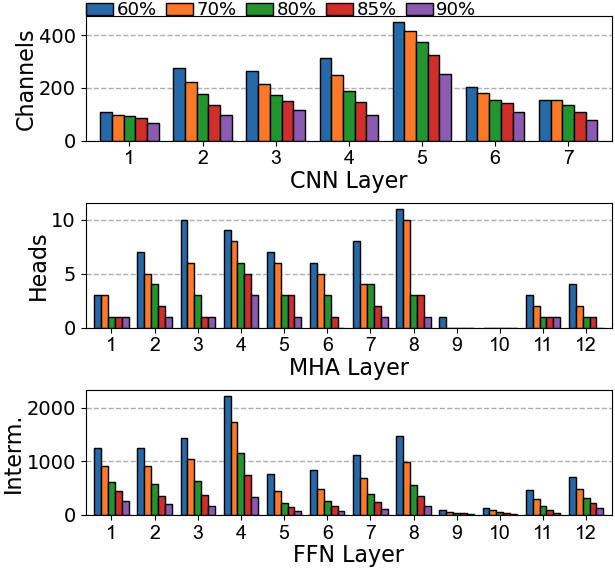}
  \caption{Visualizations of CNN channels, attention heads, and FFN intermediate dimensions from the pruned WavLM Base+. The original sizes of these components are 512, 12, and 3072, respectively. The pruning sparsity levels are set to \{60\%, 70\%, 80\%, 85\%, 90\%\}.}
  \label{fig:visualization}
  \vspace{-0.2cm}
\end{figure}

Figure~\ref{fig:visualization} shows how the number of parameters retained in the convolutional (CNN), multi-head attention (MHA), and feed-forward network (FFN) components of WavLM Base+ changes under different sparsity levels. The middle CNN layers are more strongly preserved, indicating higher task relevance. Among the Transformer blocks, the 9th and 10th layers are almost completely removed after pruning. However, due to residual connections~\cite{he2016deep}, information can still propagate to deeper layers even when attention heads are removed. 

In most prunable units, higher sparsity yields a roughly proportional reduction in parameters. However, for some components---such as the first and last two MHA layers---the structure remains unchanged beyond a certain sparsity level, suggesting that the current pruning strategy has reached its structural limit for these parts.

\subsection{Generalization beyond meeting scenarios}
\label{sec: compound_datasets}
As discussed above, we have demonstrated that our proposed pruning method can safely remove over 80\% of redundant parameters while maintaining performance comparable to the unpruned model. However, since the training data consist solely of meeting recordings from AMI, AISHELL-4, and AliMeeting, our analysis is limited to the meeting scenario and may not generalize well to other domains such as phone calls, YouTube recordings, or daily conversations.

Our ambition is to develop a general diarization pipeline, based on EEND-VC, that is both simple and effective across diverse scenarios. An ideal approach is expected to have the following characteristics:  
1) the EEND model is lightweight and easy to deploy;  
2) the clustering component uses the same hyperparameters across datasets, avoiding over-tuning for each individual case; and  
3) the overall pipeline achieves competitive performance compared to state-of-the-art systems.  

Following these principles, we pre-trained WavLM-based EEND models on a larger, multi-domain compound dataset comprising AMI, AISHELL-4 (AIS-4), AliMeeting (AliM), NOTSOFAR-1 (NSF), MSDWild (MSD), DIHARD-3 (DH3), RAMC, and VoxConverse (VoxC). Further details about these datasets are provided in Section~\ref{subsec: datasets}. We then applied our pruning strategy to the pre-trained models to eliminate redundant parameters. For speaker-embedding clustering, we adopt VBx~\cite{landini2022bayesian}, which demonstrates superior performance in our experimental setup. A detailed analysis of VBx clustering within the DiariZen framework can be found in~\cite{vbx_petr}.

The results are presented in Table~\ref{tab:compound_results}, where we also examine the effect of using longer \emph{local EEND windows} (i.e., the segment length in the EEND-VC framework). Using longer windows increase the likelihood that each speaker is captured with more speech, leading to more representative and discriminative speaker embeddings. However, they also increase the chance that more speakers co-occur within the same window, which can conflict with the assumptions of the powerset formulation---by default trained with up to four speakers and at most two overlapping. Such violations can degrade local EEND performance.

In our experiments, the local window size during training for both the unpruned WavLM Base+ and WavLM Large models, as well as for the distillation-guided pruning stage, was fixed at 8 seconds.
During inference, window lengths of 8 or 12 seconds were used. Increasing the window length during training is challenging due to GPU memory requirements; however, once WavLM is pruned, the reduced model size makes longer-window fine-tuning feasible. We therefore experimented with 12- and 16-second windows for both fine-tuning and inference of the pruned model, while longer windows exceeded our available GPU memory capacity.

As shown in the table, longer local windows generally improve performance. However, the default powerset configuration assumes up to four speakers with at most two overlapping speakers, which may not always hold in practice. To examine this effect, we adjust the overlapping constraint in the powerset~\cite{plaquet2023powerset} configuration to allow up to four overlapping speakers. This relaxed assumption significantly reduces DER---primarily the MISS error---on the AliMeeting and NOTSOFAR-1 datasets. Notably, for pruned WavLM Large, our best setup achieves superior performance to current state-of-the-art systems on most datasets. It is important to note, however, that most state-of-the-art results are obtained through extensive fine-tuning on individual datasets. In contrast, our systems are trained without any further domain adaptation and use identical hyperparameters across all datasets. As suggested by prior works~\cite{plaquet2023powerset, plaquet2024mambabasedsegmentationmodelspeaker}, although dataset-specific tuning can lead to better performance, it contradicts our objective of building a generalizable diarization system.

Another noteworthy observation is that WavLM Base+ and WavLM Large exhibit different behaviors under pruning. At 80\% sparsity, compared to the unpruned model, the pruned Base+ model performs worse across all datasets (see the 1st and 3rd rows), whereas the pruned WavLM Large generally shows improved performance (see the 7th and 8th rows). This suggests that pruning may act as a form of regularization for larger models, alleviating overfitting and leading to better generalization across diverse domains. These findings point to the potential of applying our method to other large-model domains, such as large language models, where high parameter counts have increasingly become the norm.

% \begin{table}[tbp]
% \centering
% \begin{threeparttable}
%   \caption{Out-of-domain performance on CHiME-6. Segment length used for both training and inference.}
%   \setlength{\tabcolsep}{1.8mm}
%   \label{tab:ood_chime}
%   \begin{tabular}{l |cc |cc | cc}
%     \hline
%         \multirow{2}{*}{System} & \multirow{2}{*}{Sparsity} & \multirow{2}{*}{Segment} & 
%         \multicolumn{2}{c|}{collar=0s} &
%         \multicolumn{2}{c}{collar=0.25s} \\
%       &  & & Dev  & Eval  & Dev & Eval\\
%       \hline
%       \multirow{3}{*}{WavLM Base+} & 0\% & 8s & 39.4 & 44.9 & 34.5 & 38.4 \\
%             & 80\% & 8s & 40.3 & 44.3 & 35.6 & 37.8 \\
%             & 80\% & 16s & 39.9 & 44.0 & 35.1 & 37.5 \\
%       \hline
%       \multirow{4}{*}{WavLM Large} & 0\% & 8s & 40.9 & 42.4 & 35.8 & 35.4 \\
%             & 80\% & 8s & 40.9 & 43.3 & 35.8 & 37.0 \\
%             & 80\% & 16s & 41.2 & 42.0 & 36.2 & 35.4 \\
%     \qquad w/ 4spk overlap pwrs.       & 80\% & 16s & 38.0 & 40.0 & 32.8 & 33.3 \\
%      \qquad \qquad +DOVER-Lap\tnote{*}       & 80\% & 16s & 38.0 & 40.0 & 32.8 & 33.3 \\
%     \hline
%   \end{tabular}
%   \begin{tablenotes}
%     \footnotesize
%     \item[*] The DOVER-Lap\cite{raj2021dover} across arrays is applied.
%     \end{tablenotes}
%   \end{threeparttable}
% \end{table}

\subsection{Out-of-domain evaluation on CHiME-6}
\label{subsec: ood_chime}
While our model achieves strong performance on in-domain datasets, an important question arises: How well does it generalize to out-of-domain (OOD) scenarios? On one hand, model pruning substantially reduces the number of parameters, which could increase sensitivity to domain shifts. On the other hand, the pruning mechanism in our framework introduces a regularization effect that may enhance robustness. By removing redundant parameters, the model is encouraged to learn more intrinsic and transferable representations, which are essential for maintaining performance under unseen conditions.

To investigate the out-of-domain generalization capability of our approach, we evaluate the pre-trained models listed in Table~\ref{tab:compound_results}, specifically rows~1, 3, and 6 for WavLM Base+, and rows~7, 8, and 10 for WavLM Large, on the CHiME-6 dataset~\cite{watanabe2020chime}. The results are summarized in Table~\ref{tab:ood_chime}.  
Since CHiME-6 consists of recordings from six microphone arrays, each equipped with four channels, we follow the official Kaldi recipe~\cite{povey2011kaldi} for preprocessing, including Weighted Prediction Error (WPE)~\cite{drude2018nara} dereverberation and BeamformIt~\cite{anguera2007acoustic} beamforming. The beamformed audio from the first array is used. 
% for diarization evaluation. 

As shown in the results, both the WavLM Base+ and WavLM Large pruned models achieve strong performance in this out-of-domain setting. While pruning does not always yield further gains, it provides up to 80\% parameter reduction with only minor degradation. Moreover, using longer input windows leads to consistently better results, consistent with the trends observed in Table~\ref{tab:compound_results}. These findings indicate that our pruning method remains robust under domain shifts and can be effectively applied to unseen real-world scenarios.

In the last row of Table~\ref{tab:ood_chime}, we apply DOVER-Lap~\cite{raj2021dover} to fuse the diarization outputs obtained from the beamformed recordings of each microphone array. Without any domain adaptation, our system achieves a DER of 33.3\% (with a 0.25\,s collar) on the CHiME-6 evaluation set, which is comparable to the third-place system in the CHiME-7 challenge\footnote{\url{https://www.chimechallenge.org/challenges/chime7/task1/results}}.  
It should be noted that applying BeamformIt independently to each device is not permitted under the official CHiME-7 rules. Nevertheless, our objective here is to analyze the out-of-domain generalization capability of our method rather than to compete within the challenge framework. Overall, these results demonstrate that the pruned models maintain competitive performance on in-domain datasets and generalize effectively to challenging out-of-domain conditions.

\begin{table}[tbp]
\centering
\begin{threeparttable}
  \caption{Out-of-domain performance on CHiME-6. Segment length used for both training and inference.}
  \setlength{\tabcolsep}{1.0mm}
  \label{tab:ood_chime}
  \begin{tabular}{l |cc |cc | cc}
    \hline
        \multirow{2}{*}{System} & \multirow{2}{*}{Sparsity} & \multirow{2}{*}{Window} & 
        \multicolumn{2}{c|}{collar=0s} &
        \multicolumn{2}{c}{collar=0.25s} \\
      &  & & Dev  & Eval  & Dev & Eval\\
      \hline
      \multirow{3}{*}{WavLM Base+} & 0\% & 8s & 39.4 & 44.9 & 34.5 & 38.4 \\
            & 80\% & 8s & 40.3 & 44.3 & 35.6 & 37.8 \\
            & 80\% & 16s & 39.9 & 44.0 & 35.1 & 37.5 \\
      \hline
      \multirow{4}{*}{WavLM Large} & 0\% & 8s & 40.9 & 42.4 & 35.8 & 35.4 \\
            & 80\% & 8s & 40.9 & 43.3 & 35.8 & 37.0 \\
            & 80\% & 16s & 41.2 & 42.0 & 36.2 & 35.4 \\
     \qquad + DOVER-Lap\tnote{*}   & 80\% & 16s & 38.0 & 40.0 & 32.8 & 33.3 \\
    \hline
  \end{tabular}
  \begin{tablenotes}
    \footnotesize
    \item[*] The DOVER-Lap\cite{raj2021dover} across arrays is applied.
    \end{tablenotes}
  \end{threeparttable}
  \vspace{-0.3cm}
\end{table}

\section{Conclusion}
In this paper, we presented an in-depth study on compressing the self-supervised WavLM model for end-to-end speaker diarization through structured pruning. Our approach achieves up to 80\% model size reduction and 4× inference acceleration without compromising performance. We explored multiple pruning strategies, including MAC-based objectives, component-wise pruning, and progressive sparsity schedules. Furthermore, we showed that pruning with reduced data, followed by fine-tuning on the full dataset, offers an efficient training strategy that preserves performance while substantially reducing computational cost. Extensive evaluations across eight diverse diarization datasets demonstrate that the pruned models not only retain state-of-the-art performance in-domain but also generalize effectively to out-of-domain conditions, including strong results on the CHiME-6 dataset without domain adaptation. All models and code are publicly released to support reproducibility and further research in speaker diarization and model compression.

\section{Acknowledgements}
% The work was supported by Ministry of Education, Youth and Sports of the Czech Republic (MoE) through the OP JAK project "Linguistics, Artificial Intelligence and Language and Speech Technologies: from Research to Applications" (ID:CZ.02.01.01/00/23\_020/0008518). Computing on IT4I supercomputer was supported by MoE through the e-INFRA CZ (ID:90254).
The work was supported by Ministry of Education, Youth and Sports of the Czech Republic (MoE) through the OP JAK project "Linguistics, Artificial Intelligence and Language and Speech Technologies: from Research to Applications" (ID:CZ.02.01.01/00/23\_020/0008518). Computing on IT4I supercomputer was supported by MoE through the e-INFRA CZ (ID:90254).

\bibliographystyle{IEEEtran}
\bibliography{mybib}

\end{document}